\documentclass[twocolumn,amsmath,amssymb]{revtex4}
\usepackage{dcolumn}
\usepackage{bm}

\begin{document}


\title{Comment on "Parity Doubling and $\mathbf{SU(2)_L\times SU(2)_R}$ 
Restoration in the Hadronic Spectrum" and "Parity Doubling Among the Baryons"}

\maketitle

In a recent paper~\cite{ja} the authors arrived at the following result:
{\it In a world with spontaneous symmetry
breaking and massless pions, the chiral symmetry realized in a
linear way gives no relations among the properties of hadron
states, such as masses and couplings. Such predictions would hold
only if certain chirally invariant operators are dynamically
suppressed.} The final conclusion was that restoration of 
$SU(2)_L\times SU(2)_R$ cannot occur in the presence of massless pions
at zero temperature and chemical potential. Needless to say that 
such a conclusion, if true, has a fundamental character. Subsequently 
Comment~\cite{Gl} appeared where it was emphasized that paper~\cite{ja}
"may create a false impression about the nature of the problem" and that
suppression of offending operators occurs naturally high in the spectrum.
While we agree with the conclusion about "false impression", we did not find
in~\cite{Gl} a clear-cut explanation why the suppression of chirally 
invariant operators must be natural (in the cited papers some general 
quasiclassical arguments were used to justify the chiral symmetry 
restoration, but how the quasiclassics suppresses the chirally invariant
operators of~\cite{ja} was not shown). We would like to comment on the 
result of~\cite{ja} using another arguments.

It has been known for long ago (see, e.g.,~\cite{wein}) that non-linearly 
realized chiral symmetry (i.e. the Nambu-Goldstone realization) is a dynamic 
symmetry. Unlike algebraic symmetries, as, e.g., the linearly realized chiral
symmetry (the Wigner-Weil realization), dynamic symmetries alone do not lead to
any
algebraic relations. The result of~\cite{ja} is just a demonstration of this
fact in the language of effective field theory. If one applies such a 
description to high hadron excitations, of course, one inevitably arrives at 
the conclusion in~\cite{ja}, nothing surprising. As was stressed in the same
seminal paper~\cite{wein}, algebraic consequences of non-linearly realized
chiral symmetry follow from restrictions on asymptotic behavior of amplitudes
at high energy. The famous Weinberg's sum rules~\cite{wein2} and resulting 
(using KSFR relation) formula
$m_{a_1}=\sqrt{2}\,m_{\rho}$ can be considered as classical examples of such
relations, which were derived from asymptotic restrictions on high energy 
behavior of two-point correlators of vector (V) and axial (A) currents. 
All this was known before QCD. Now we know from the asymptotic freedom of QCD
that these restrictions are governed by the perturbation theory, i.e. they
are chirally symmetric in a usual algebraic sense. If one saturates the 
V,A correlators by "one resonance $+$ perturbative continuum" one obtains
different masses of states because pion contributes significantly  to the 
A-channel due to PCAC. But the radial excitations of $a_1$-meson do not
have this contribution due to generalized PCAC. To correctly define the high
meson excitations one needs the large-$N_c$ limit. The number of excitations 
then must be infinite in order to reproduce the perturbative asymptotics of 
correlators. Thus, saturating the correlators by infinite number of resonances,
the relative role of pion contribution, in a sense, becomes infinitely small.
Since the resonances reproduce the perturbative asymptotics, the spectrum
should "repeat" the chiral invariance of perturbation theory. In this sense the 
algebraic chiral
symmetry is restored in the spectrum of radial excitations. Generalization
of this idea to arbitrary channels is straightforward. To our knowledge,
first it was proposed in~\cite{AV}. Subsequently it was widely used in
different models (see, e.g.,~\cite{AV2}) and, moreover, it was directly 
derived in various papers within QCD sum rules~\cite{we}. In some
sum rule analyses~\cite{others} the (masses)$^2$ of chiral partners have
a constant shift at any energy. In this case one has an effective restoration
of algebraic chiral symmetry in the sense that the corresponding
masses tend to degeneracy (since they grow), but with a slower rate~\cite{we2}.
Even in that case there are algebraic relations between hadronic parameters.

Thus, finally one has the asymptotically degenerate mass spectrum for chiral 
partners. The situation is indistinguishable from Wigner-Weyl realization
of chiral symmetry. Although the large-$N_c$ limit usually works quite well,
the argument might seem not compelling because in the real world $N_c=3$.
But what should be emphasized here is that such a phenomenon is expected 
to occur due to the asymptotic freedom of QCD, the latter was completely 
ignored in analysis~\cite{ja}. Of course, one could speculate that it is 
asymptotic freedom which is responsible for the dynamical suppression of 
chirally invariant operators. But for us it looks more likely that the very 
problem of "chirally invariant operators" is in doubt. Low-energy pion-like 
effective Lagrangians (like those used in Ref.~\cite{ja}) reflect only some 
features of QCD, their application is restricted and, hence, they seem not to
be 
trustworthy for drawing very general statements which are valid at any energy. 

Let us give a simple example. Let $\varphi_+$ and $\varphi_-$ be fields of
chiral 
partners having masses $m_+$ and $m_-$ correspondingly. In Lagrangian~(9) 
of Ref.~\cite{ja} the term breaking mass degeneracy is proportional to the
factor $m_1$ and
it describes different reactions with pions, e.g., the decay
$\varphi_+\rightarrow\varphi_-\pi$ 
(we assume $m_+>m_-$). Thus, such a language is relevant to the reality if this
decay 
indeed can occur (together with decays like
$\varphi_+\rightarrow\varphi_-\pi\pi\pi$). 
This works well for the ground states. Let us consider, however, the 
radial excitations. Denote $m_+^2(n)-m_-^2(n)\equiv\delta(n)$ ($n$ is the
radial 
quantum number). If $\delta(n)$ is not decreasing one has no genuine restoration
of algebraic chiral symmetry. Even if we assume this variant, 
in the real world the masses grow 
with $n$ rapidly enough to ensure decreasing 
$m_+(n)-m_-(n)=\delta(n)/(m_+(n)+m_-(n))$ such that one usually has 
$m_+(n)-m_-(n)\lesssim m_{\pi}$ at $n\gtrsim3$, i.e. this decay is impossible
any
more since some excitation. The conclusions for the real world inferred from the 
effective Lagrangian become just artefacts 
of chiral limit. On the other hand, this limit is expected to be a very good
first
approximation to the reality. To meet this expectation one should take $m_1=0$ 
for high enough resonances. But the chiral partners are then degenerate. 
The same can be repeated for the part involving
the covariant derivative of the pion field, Eq.~(12) in Ref.~\cite{ja}. 

The authors of Ref.~\cite{ja} argued in a companion paper~\cite{ja2} that
inclusion
of finite width (i.e. taking into account next-to-leading order in the
large-$N_c$
counting) into QCD sum rules can, in principle, destroy the chiral symmetry 
restoration. Skipping some possible technical problems, we would like to note 
that unlike the string-like mass spectrum for light mesons, $m^2(n)\sim n$, 
the string-motivated formula $\Gamma(n)\sim \sqrt{n}$ is not
supported experimentally. This can be straightforwardly checked by looking at
Particle Data~\cite{pdg} and review~\cite{bugg}. Of course, may be this
relation 
sets in at a very 
large $n$, like in the 't Hooft model (planar dim2 QCD), nevertheless there is
no much belief even in this hypothesis. A reason can be easily indicated:
Many high radial and spin meson excitations prefer to decay into three or four
particles in contrast to a naive string-based picture predicting the
dominance of two-particle decay. It means also that the large-$N_c$ counting 
fails to work for $\Gamma(n)$. To our knowledge, there is no clear 
explanation of this phenomenon in the literature. For this reason at present
stage any similar considerations of finite widths look highly speculative. 

In Ref.~\cite{ja2} the analysis carried out in~\cite{wein} was taken as an
example
of advocated ideology. We would like to remind that Ref.~\cite{wein} deals with
the ground states only. A result is that at any given helicity the mass matrix
$m^2$ may be written as the sum of a chiral scalar $m_0^2$ and the fourth
component
of a chiral four-vector $m_4^2$ with respect to $SU(2)_L\times SU(2)_R$ formed
by
the isospin $\mathbf T$ and the axial vector coupling matrix $\mathbf X$. 
The term $m_4^2$ appears due to existence of vacuum Regge trajectories and
destroys
the algebraic chiral symmetry. In order to extrapolate this conclusion to
higher excitations, evidently, one has to apply the logic used in~\cite{wein}
to these states. However, from experiment (a strong suppression of reactions
with
one pion for many higher meson states) and generalized PCAC the matrix $\mathbf
X$
is expected to be trivial resulting in $m_4^2=0$ (from Eq.~(1.14) in
Ref.~\cite{wein}).
This observation is only suggestive, just demonstrating that application of
results 
from Ref.~\cite{wein} to higher excitations needs a serious reanalysis.

\vspace{0.3cm}

\noindent
S. S. Afonin\\
Dep. d'Estructura i Constituents de la Ma\-t\`e\-ria and
CER for Astrophysics, Particle 
Physics and Cosmology,
Universitat de Barcelona, 647 Diagonal, 08028, Spain


\begin{thebibliography}{99}
\bibitem{ja} R. L. Jaffe, D. Pirjol, and A. Scardicchio, Phys. Rev. Lett.~{\bf
96}, 
121601 (2006) [hep-ph/0511081].
\bibitem{Gl} T. D. Cohen and L. Ya. Glozman, hep-ph/0603240.
\bibitem{wein} S. Weinberg, Phys. Rev. {\bf 177}, 2604 (1969).
\bibitem{wein2} S. Weinberg, Phys. Rev. Lett. {\bf 18}, 507 (1967).
\bibitem{AV} A.~A.~Andrianov and V.~A.~Andrianov, hep-ph/9705364
[Zapiski Nauch. Sem. POMI {\bf 245}, 5 (1996)].
\bibitem{AV2} A. A.~Andrianov, D. Espriu, and R. Tarrach, Nucl. Phys. B {\bf
533}, 429 (1998);
A.~A.~Andrianov and D.~Espriu, JHEP {\bf 10}, 022 (1999);
A.~A.~Andrianov and V.~A.~Andrianov, hep-ph/9911383;
A.~A.~Andrianov, V.~A.~Andrianov, and S.~S.~Afonin, hep-ph/0101245;
hep-ph/0209260;
V.~A.~Andrianov and S.~S.~Afonin, Eur. Phys. J. A {\bf 17}, 111
(2003); J. Math. Sc. {\bf 125}, 99 (2005) [Zapiski Nauch. Sem.
POMI {\bf 291}, 5 (2002)]; hep-ph/0304140; Phys. At. Nucl. {\bf
65}, 1862 (2002) [Yad. Fiz. {\bf 65}, 1913 (2002)].
\bibitem{we} S.~R.~Beane, Phys. Rev. D~{\bf 64}, 116010 (2001); 
S. S.~Afonin, A. A.~Andrianov, V. A.~Andrianov, and D.~Espriu,
JHEP {\bf 04}, 039 (2004) [hep-ph/0403268]; hep-ph/0509144; 
M. Shifman, hep-ph/0507246; S. S.~Afonin and D.~Espriu, hep-ph/0602219.
\bibitem{others} E.~R.~Arriola and W.~Broniowski, hep-ph/0603263;
O. Cat\`a, M. Golterman, and S. Peris, hep-ph/0602194;
S. S. Afonin, Phys. Lett. B {\bf 576}, 122 (2003);
A.~A.~Andrianov, V.~A.~Andrianov, and S.~S.~Afonin,
hep-ph/0212171.
\bibitem{we2} S.~S.~Afonin, hep-ph/0603166.
\bibitem{ja2} R. L. Jaffe, D. Pirjol, and A. Scardicchio, hep-ph/0602010.
\bibitem{pdg} S. Eidelman {\it et al.}, Phys. Lett. B~{\bf 592}, 1 (2004).
\bibitem{bugg} D. V. Bugg, Phys. Rept. {\bf 397}, 257 (2004).
\end{thebibliography}
\end{document}